\documentclass[journal]{IEEEtran}
\usepackage{amsmath,amssymb,amsfonts}
\usepackage{graphicx}
\usepackage{booktabs}
\usepackage{multirow}
\usepackage{tabularx}
\usepackage{array}
\usepackage{xcolor}
\usepackage{tikz}
\usetikzlibrary{positioning, arrows.meta, calc, fit, backgrounds, decorations.pathreplacing}
\usepackage{pgfplots}
\usepgfplotslibrary{groupplots}
\pgfplotsset{compat=1.18}
\usepackage{cite}
\usepackage{url}
\usepackage[hidelinks]{hyperref}

\newcolumntype{Y}{>{\raggedright\arraybackslash}X}

\usepackage{enumitem}
\setlist[itemize]{topsep=1pt, itemsep=0.5pt, parsep=0pt, leftmargin=1.2em}
\AtBeginDocument{%
\setlength{\abovedisplayskip}{4pt plus 1pt minus 1pt}%
\setlength{\belowdisplayskip}{4pt plus 1pt minus 1pt}%
\setlength{\abovedisplayshortskip}{2pt}%
\setlength{\belowdisplayshortskip}{2pt}}

\begin{document}

\title{Semantic Error Control Coding with Foundation Models for Future Communications}

\author{Chentao~Yue,~\IEEEmembership{Senior Member,~IEEE,}
        Gaoyang~Pang,~\IEEEmembership{Member,~IEEE,}
        Branka~Vucetic,~\IEEEmembership{Life Fellow,~IEEE,}
        and~Yonghui~Li,~\IEEEmembership{Fellow,~IEEE}%
\thanks{The authors are with the School of Electrical and Computer Engineering, The University of Sydney, Sydney, NSW 2006, Australia. Email: \{chentao.yue, gaoyang.pang, branka.vucetic, yonghui.li\}@sydney.edu.au.}%
\thanks{This work was supported by ARC DECRA under Grant DE250101332.}%
}

\maketitle

\begin{abstract}
Classical channel decoding typically treats all information sequences as equally likely and relies primarily on the channel observations and code structure, without exploiting statistical or semantic structure in the source data. Although source compression is designed to remove redundancy, practical source coding can leave substantial residual structure that conventional channel decoders do not exploit. Modern multimodal data sources, including text, speech, and images, exhibit rich statistical and semantic dependencies that foundation models can learn and exploit to improve channel decoding. This article introduces semantic error control coding (SECC), which seamlessly integrates the semantic structure of the source into encoding and decoding through a foundation model. The semantic source prior, represented by the model's a priori probability of the source content, directs code redundancy toward semantically important content at the encoder, and improves reliability estimation, candidate search, and error detection/correction at the decoder. The channel code keeps its algebraic structure, and its constraints ensure that the semantic suggestions from the foundation model comply with this structure. We describe the SECC framework, classify its design methods into three approaches, and demonstrate representative schemes on text sources. The demonstrated schemes show several decibels of coding gain over conventional decoding on AWGN channels, and reach error rates below the normal approximation bound. Finally, we identify several open challenges.
\end{abstract}


\vspace{-10pt}
\section{Introduction}
\vspace{-3pt}
Modern communication systems are built on the separation between source coding and channel coding. The source encoder removes redundancy, the channel encoder adds structured redundancy, and the channel decoder recovers the transmitted bits from noisy observations. This architecture, established by Shannon's information theory \cite{Shannon1948}, is analyzable, modular, and widely deployed in wireless standards.

These systems, however, increasingly carry structured source data (text, speech, images, video, control messages) that are far from random bit strings. Their syntax, context, and correlations across symbols, packets, and modalities create redundancy that a receiver can exploit alongside channel observations. This motivates semantic communications, which use source meaning to improve transmission reliability \cite{Gunduz2023BeyondBits}.

A prominent approach is deep joint source-channel coding (JSCC), where a neural network maps the source directly to channel symbols and jointly learns compression and protection, with strong results for images and text \cite{Bourtsoulatze2019}. Compression and protection are carried out by the same network, so neither can be separated, replaced, or verified independently. Adapting to new channels or data rates may require retraining the entire system. Without an explicit error-control layer, reliability is provided by the learned model rather than enforced by code constraints, and no independent structure verifies the decoder output. Token communication systems \cite{Qiao2025TokCom} share this limitation. They transmit foundation-model tokens and recover corrupted content with generative models.

Another stream of work keeps the channel code unchanged and makes the decoder source-aware. Source-controlled channel decoding biased decoder decisions with Markov chains and symbol-frequency tables \cite{Hagenauer1995}, but was limited to analytically tractable distributions over short sequences. Foundation models, including large language models (LLMs) and vision-language models (VLMs), remove this limitation. They can estimate complex, long-range source structure and provide source likelihoods, confidence scores, or candidate reconstructions based on contextual information. Recent studies combine these outputs with the channel observations when scoring decoder candidates, and code constraints keep the output a valid codeword \cite{Hao2026SemOSD,Li2026LLMViterbi}. In principle, the same priors can also guide redundancy allocation at the transmitter. A systematic framework that defines the interaction between the model and the coding chain, however, is still missing. 

This article proposes \emph{semantic error control coding} (SECC), which systematically exploits the semantic structure of the source in encoding and decoding. Unlike JSCC, which learns a new source-to-channel representation, SECC preserves the code structure, and a foundation model supplies source-prior information that complements the channel code. We describe the SECC framework, where the transmitter keeps a standard channel code and may reallocate its redundancy by semantic importance, while the receiver integrates its channel decoder and a foundation model into a semantic decoder that exploits residual source redundancy. We classify the design methods into three approaches, \emph{inference helps coding}, \emph{coding helps inference}, and \emph{mutual reinforcement}. We then introduce representative SECC schemes on text sources, demonstrating coding gains of several decibels over conventional decoding. Finally, we identify open challenges in encoder-side design, performance limits, standard compatibility, and the trade-off between model size and coding gain. SECC particularly benefits power-limited links that tolerate processing delay, such as deep-space links with round-trip times of minutes. Emergency messaging and machine-type uplinks benefit through their short structured text and telemetry. Latency-critical applications become viable only with faster inference, from hardware progress or lightweight models.

\vspace{-6pt}
\section{Codes, Sources, and Foundation Models}
\label{sec:framework}

\subsection{Classical Channel Coding}

Consider a binary linear code $\mathcal{C}(n,k)$ that maps $k$ information bits $\mathbf{m} \in \{0,1\}^{k}$ to an $n$-bit codeword $\mathbf{c} \in \{0,1\}^{n}$ at rate $R = k/n$. After modulation and transmission over a noisy channel, the receiver observes $\mathbf{y}$ and seeks the most likely transmitted codeword. Maximum-likelihood decoding (MLD) solves
\begin{equation}
\hat{\mathbf{c}}_{\mathrm{ML}} = \arg\max_{\mathbf{c} \in \mathcal{C}} \, P(\mathbf{y} \mid \mathbf{c}),
\label{eq:mld}
\end{equation}
which treats all $2^{k}$ information sequences as equally probable.

This assumption of a \emph{uniform source}, i.e., equiprobable information sequences, underpins channel coding theory. Shannon's channel coding theorem establishes the capacity of a memoryless channel under uniform input \cite{Shannon1948}. Finite-blocklength bounds then characterize the best achievable block error rate (BLER) for a given blocklength and rate~\cite{Polyanskiy2010}. 

The assumption simplifies the decoder design for different channel codes, including Viterbi, belief propagation (BP), successive cancellation list (SCL), or ordered-statistics decoding (OSD) \cite{Yue2022CommMag}. Once a source has been compressed into bits, the channel decoder does not need to know what the bits represent, and any residual structure in the information sequence is ignored.

\vspace{-7pt}
\subsection{Source Coding and Semantic Priors}

Source coding for transmission takes one of two forms, each leaving a different amount of structure for the SECC receiver.

\begin{itemize}
\item \emph{Conventional compression}. Huffman and arithmetic coding compress the source toward the entropy rate, producing a near-uniform, memoryless bit stream for which MLD is optimal. Variable-length codewords lack synchronization boundaries, so a residual bit error propagates through the source decoder and can corrupt the entire message. Still, one usable property survives. A corrupted stream decompresses into garbled output, so the receiver, after source decoding, can still judge whether the whole message is plausible. Inside the compressed stream itself, no finer structure remains for the channel decoder to exploit.
\item \emph{Tokenization}. Subword tokenizers, such as byte-pair encoding (BPE) and WordPiece, map frequent character patterns of varying length to single tokens, reducing sequence length while retaining statistical dependencies. Each token, regardless of its character span, maps to a fixed-length binary index, so a bit error corrupts only the token that contains it. In SECC, tokenization can serve as a fixed-length source code, and the channel decoder exploits the redundancy it retains.
\end{itemize}

Compared with conventional compression, tokenization pays rate for structure. A pretrained language model can estimate the conditional distribution of each token, yielding a \emph{source prior} $P(\mathbf{m} \mid \mathcal{S})$ on the message $\mathbf{m}$ given context $\mathcal{S}$. This prior concentrates probability mass on plausible messages, reducing the channel decoder's search space.

The rate cost depends on the modality. For text, a BPE vocabulary of 50{,}000 subword tokens requires 16 bits to represent each token. With an average of four characters per token, the rate is about 4 bits per character, roughly 40 percent more than DEFLATE, the compressor standardized for uplink data compression in 3GPP NR. Both remain above the 0.6 to 1.3 bits per character entropy of printed English \cite{Shannon1951}. For images and video, learned tokenizers such as MAGVIT-v2 match HEVC and VVC in perceptual quality near 0.04 bits per pixel \cite{Yu2024MAGVIT}. The cost lies in reconstruction rather than rate. The detokenizer is generative, so its output cannot be exact at the pixel level. The measured coding gain from the semantic prior must outweigh the rate cost of forgoing compression, a trade-off quantified in Section~\ref{sec:results}.

\vspace{-7pt}
\subsection{Foundation Models}

Foundation models are neural networks trained on broad corpora through self-supervision, typically by predicting the next token, and adaptable to a wide range of tasks. Next-token training performs maximum-likelihood estimation under the cross-entropy objective, so the trained model converges toward the empirical distribution of the training corpus. LLMs instantiate this estimate for text, and autoregressive multi-modal models extend it to images and video.

To a channel decoder, such a model provides three forms of information.

\begin{itemize}
    \item \emph{Likelihood estimation}. Autoregressive models such as GPT and LLaMA provide the conditional probability $P(x_t \mid x_{<t})$ of each token $x_t$ given the preceding tokens $x_{<t}$. A decoder can use these probabilities to rank candidate sequences or adjust the reliability of individual bits.
    \item \emph{Fidelity verification}. Embedding models such as Sentence-BERT \cite{Reimers2019SBERT} map an input sequence to a vector representation, and the similarity between vectors measures how well a candidate output agrees with its context. The check can operate on whole messages, so it stays compatible with conventional compression, applied after the message is decompressed.
    \item \emph{Reconstruction}. Generative models trained or prompted to fill in missing spans, such as BART \cite{Lewis2020BART}, regenerate corrupted segments from surrounding context.
\end{itemize}

 Each output requires a full forward pass through the model, so a receiver must count model calls as part of its decoding cost.

\vspace{-6pt}
\section{The Semantic Error Control Coding Framework}
\label{sec:secc}

\definecolor{txcolor}{RGB}{47,74,112}
\definecolor{rxcolor}{RGB}{47,74,112}
\definecolor{fmcolor}{RGB}{122,49,64}
\definecolor{chcolor}{RGB}{120,120,120}
\definecolor{anncolor}{RGB}{100,100,100}

\begin{figure*}[t]
\centering
\includegraphics{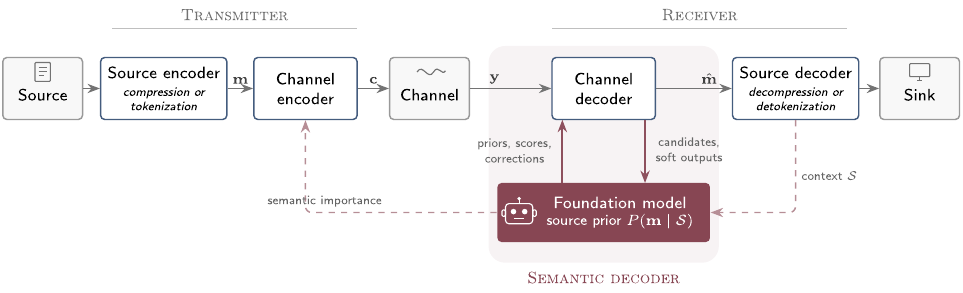}
\caption{The SECC framework. The channel decoder and the foundation model form the semantic decoder (shaded), where the model supplies the source prior $P(\mathbf{m}\mid\mathcal{S})$ and the code constraints keep the output a valid codeword. Dashed paths are optional. }
\label{fig:architecture}
\end{figure*}

\subsection{Framework and Performance Metrics}

Fig.~\ref{fig:architecture} illustrates the SECC framework. The transmitter compresses or tokenizes the source messages and encodes the result with a standard channel code (LDPC, polar, convolutional, or other structured codes). At the receiver, the channel decoder and the foundation model form an integrated \emph{semantic decoder}. The foundation model supplies source priors, reliability estimates, or candidate scores that assist the decoder's search. Code constraints such as parity checks keep the decoder output a valid codeword. The model's context may come from tentative decoding results of the same message, from previous messages, or from known structure such as message formats. Semantic knowledge enters the coding chain at three points: redundancy allocation at the encoder, prior-guided decoding, and meaning-level evaluation at the decoder output. Encoder-side use assumes that the transmitter holds a copy of the same model, so both ends compute the same prior.

The performance of SECC is judged on three levels.

\begin{itemize}
\item \emph{Block-level reliability}. BLER counts decoding failures and is directly comparable to conventional channel coding and finite-blocklength benchmarks.
\item \emph{Meaning-level fidelity}. Unlike BLER, these metrics depend on the source. For text, the cosine similarity of Sentence-BERT (SBERT) embeddings compares meaning directly \cite{Reimers2019SBERT}. For images and video, PSNR measures pixel-level fidelity, and SSIM and LPIPS measure structural and perceptual similarity.
\item \emph{Task-level performance}. When the receiver serves a specific task, the task loss itself is the final criterion.
\end{itemize}

A single bit error can invalidate a control command, while a sentence differing in several tokens can preserve its meaning. Bit-level correctness and preserved meaning are therefore not equivalent, so on structured sources the three levels must be evaluated separately.

\begin{figure*}[t]
\centering
\includegraphics[width=\linewidth]{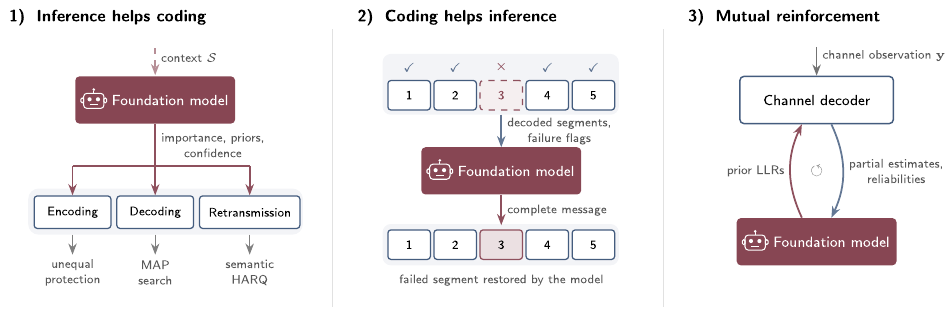}
\caption{The three design approaches. 1) Model outputs guide encoding, decoding, and retransmission. 2) Segments are protected by short codes; the code flags the failed segment, and the model outputs the complete message. 3) The decoder and the model exchange soft information in a closed loop.}
\label{fig:approaches}
\end{figure*}

The model and the coding chain can interact in three ways. 1) \emph{Inference helps coding}, where model inference guides encoding, decoding, error detection, and retransmission. 2) \emph{Coding helps inference}, where code structure localizes and verifies model corrections. 3) \emph{Mutual reinforcement}, where the two are fused in one decoder. Fig.~\ref{fig:approaches} illustrates the three design approaches.

\vspace{-7pt}
\subsection{Approach 1: Inference Helps Coding}

At the encoder, the foundation model estimates the semantic importance of each token. A small number of tokens can determine the meaning of the entire message, as a negation or a numerical value can invert a command. Protecting these tokens more strongly therefore matters more than protecting all tokens equally. Unequal error protection (UEP) then assigns different levels of protection to different tokens. The encoder applies the puncturing and shortening patterns of standard rate matching, which adapt the number of transmitted bits to the required rate, to give important tokens more redundancy. This applies directly to the tokenized configuration, where token boundaries and their mapping to coded bits are fixed, and the receiver obtains the mapping from the shared system configuration or from control signaling. 

At the receiver, the model can supply a prior probability $P(\mathbf{m})$ for each candidate message $\mathbf{m}$, its probability under the source model. The optimal decoder combines this prior with the channel likelihood through maximum a posteriori (MAP) decoding:
\begin{equation}
\hat{\mathbf{c}}_{\mathrm{MAP}} = \arg\max_{\mathbf{c}(\mathbf{m}) \in \mathcal{C}} \, P(\mathbf{y} \mid \mathbf{c}(\mathbf{m})) \cdot P(\mathbf{m}).
\label{eq:map}
\end{equation}
The first term measures how well candidate codeword $\mathbf{c}(\mathbf{m})$ explains the received signal $\mathbf{y}$, and $P(\mathbf{m})$ measures how plausible the message is under the source model. When $P(\mathbf{m})$ is uniform, \eqref{eq:map} reduces to MLD in \eqref{eq:mld}. When it is not, two candidates that explain the received signal equally well can be separated by their source probabilities, information that the channel observations alone do not carry. Exact MAP is intractable. The search spans $2^k$ messages, and the prior does not decompose according to the code structure that fast decoders exploit. Practical schemes use model scores to rank, reweight, or prune the candidates that a classical decoder explores. For example, in semantic ordered-statistics decoding (Sem-OSD) \cite{Hao2026SemOSD}, a language model predicts each byte from the decoded context, and the predicted byte probabilities are converted into bit-level reliabilities. The decoder fuses the semantic reliabilities with the channel reliabilities and tests candidate codewords according to the fused reliability. On a non-uniform source, it outperforms conventional OSD with the same code and blocklength, and with a uniform source (no semantic structure) it reduces to conventional OSD. LLM-aided Viterbi decoding similarly uses a language model to score the decoded prefix of each surviving trellis path and to prune paths that are linguistically implausible, which is particularly useful when competing paths have similar channel metrics but differ in semantic plausibility \cite{Li2026LLMViterbi}. The same principle applies in the compression configuration, with no change at the transmitter. A list decoder generates several candidate bit streams, decompresses each, and the model ranks the resulting plaintexts by plausibility.

Inference can also strengthen error detection and retransmission. Alongside the CRC, the receiver can check semantic plausibility, such as the perplexity of the decoded text. An undetected error must then produce another valid codeword with plausible content, far rarer than passing the CRC alone. A failed semantic check triggers retransmission; semantic HARQ (SHARQ) requests additional redundancy only when decoding fails and the model cannot repair the packet from context \cite{Hao2025ShortWinsLong}.

\vspace{-7pt}
\subsection{Approach 2: Coding Helps Inference}

The complementary approach uses the channel code to assist the model in recovering corrupted content. Conventional decoding first identifies which parts of the message are reliable and which have failed. The model then reconstructs the failed parts from the reliable surrounding context.

A message can be divided into segments, each protected by its own short code and decoded in parallel. Decoding failures are identified per segment, so residual errors appear as erasures at known positions. Models such as BART are pretrained to fill masked spans from surrounding context, i.e., to repair these erasures. The short-wins-long scheme implements this segment repair \cite{Hao2025ShortWinsLong}. Segment length sets the balance between coding gain and semantic recoverability. Long segments enjoy the coding gain of large blocklength, but one failure erases much of the message and the context needed to repair it. Short segments pay the finite-blocklength penalty, but failures stay isolated within abundant context.

Generative repair can hallucinate, and code constraints can contain this risk. The receiver can ask the model for several candidate outputs, re-encode each into a codeword, and select the one closest to the received signal. In this way, plausible but wrong inference results are rejected. Semantic list decoding (SLD) follows this design \cite{Hao2025ShortWinsLong}. Since candidates are compared as complete segments, the design is compatible with both conventional compression and tokenization.

This approach can be regarded as a cross-layer design. The physical layer passes decoded segments and failure flags to the application layer, where the model operates, and the transmitter and the core decoder stay unchanged. However, the model inference never considers the channel condition or sees the channel LLRs.

\vspace{-7pt}
\subsection{Approach 3: Mutual Reinforcement}

The third approach fuses \emph{inference helps coding} and \emph{coding helps inference} in a single decoder. The residual redundancy of the source can be viewed as an implicit outer code that was never explicitly encoded, with the foundation model as its soft decoder. The channel code and the source model then provide complementary information about the transmitted message.

A natural realization is a turbo-like iterative decoder. Any soft-output decoder can participate, such as BP for LDPC codes, soft-output SCL for polar codes, or the BCJR decoder for convolutional codes. It passes partial estimates and their reliabilities to the model, and the model returns prior LLRs computed from this context, for some or all positions. The symbols decoded in one round extend the model's context, so the next prior is more accurate and the next decoding round resolves more symbols. The alternation can run as an outer loop around complete decoding attempts, or inside the decoder itself, with the prior injected every few BP iterations.

The fused decoder offers the largest potential gain because it approaches joint decoding of code and source. In turbo decoding, the component decoders exchange only extrinsic information, each subtracting what the other already supplied. A foundation model cannot perform this subtraction, so erroneous priors can propagate and amplify across decoding rounds. Practical schemes can address this issue by weighting the prior by model confidence and scaling down the injected LLRs. The model should also accept soft inputs directly. Token-level LLRs can be incorporated as probability-weighted averages of token embeddings, as speech-recognition models already do with lattices of scored candidate words.

\vspace{-6pt}
\section{Illustrative Results}
\label{sec:results}

\newcommand{\semOSD}{Sem-OSD}

Unless stated otherwise, the experiments transmit English text, the modality with the most mature source priors, using bytes as fixed-length tokens. The text consists of English sentences drawn from the SNLI and Wikipedia corpora.\footnote{Implementations of all schemes and algorithms in this section are available at \url{https://chentaoyue.github.io/semantic-error-control-coding-hub/} (accessed Aug. 2026).} We report BLER at the block level and SBERT similarity at the meaning level. SNR denotes $E_s/N_0$, the energy per channel symbol over the noise spectral density.

\vspace{-7pt}
\subsection{Inference Helps Coding: Semantic Priors in the Decoder}

\begin{figure}[t]
\centering
\includegraphics{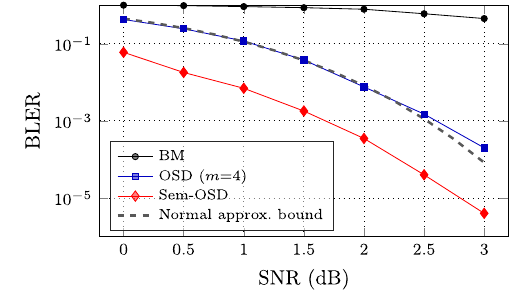}
\caption{BLER of BCH$(127,64)$ over AWGN with an English text source.}
\label{fig:bler-awgn-bch}
\end{figure}

We first evaluate semantic ordered-statistics decoding (\semOSD) on a binary BCH$(127,64)$ code over the AWGN channel \cite{Hao2026SemOSD}. Fig.~\ref{fig:bler-awgn-bch} compares three receivers. The Berlekamp--Massey (BM) algebraic decoder and conventional OSD with order $m=4$ both operate under the uniform-source assumption. \semOSD\ exploits source structure through a byte-level language model (ByT5 \cite{Xue2022ByT5}), whose byte-granularity outputs map directly to bit-level reliabilities. The normal approximation bound \cite{Polyanskiy2010} gives the best achievable BLER under uniform information. Each codeword carries only eight characters. When DEFLATE (Section~\ref{sec:framework}) is applied to blocks of this length, it outputs roughly 10 bits per character, more than the 8 bits of uncompressed ASCII. Conventional OSD over uncompressed ASCII is therefore the equal-occupancy baseline in Fig.~\ref{fig:bler-awgn-bch}.

\semOSD\ achieves a BLER below $10^{-5}$ at an SNR of 3~dB. At a BLER of $10^{-3}$, it is 0.9~dB ahead of conventional OSD, and at 3~dB its BLER is more than an order of magnitude below the normal approximation bound. The bound assumes uniform information bits, so the gap measures the coding gain available from the source redundancy. 

\vspace{-7pt}
\subsection{Coding Helps Inference: Segmentation and Semantic Recovery}

\begin{figure}[t]
\centering
\includegraphics{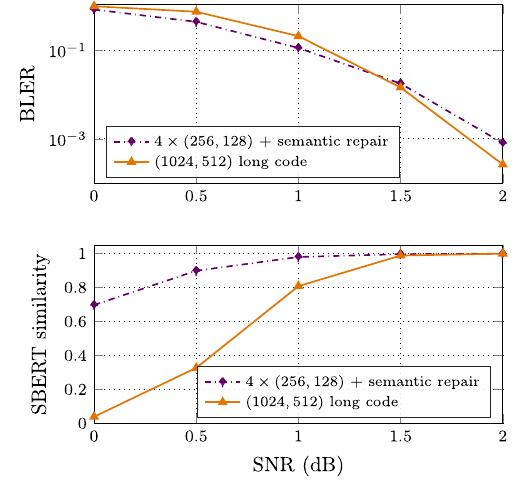}
\caption{A sentence transmitted as four $(256,128)$ coded segments with semantic repair, versus a single $(1024,512)$ long code of equal total length and rate \cite{Hao2025ShortWinsLong}.}
\label{fig:short-wins-long}
\end{figure}

Classical coding intuition favors longer blocklengths. Fig.~\ref{fig:short-wins-long} compares a single long code protecting an entire sentence with four short coded segments plus semantic repair, where the model reconstructs a failed segment from its decoded neighbors \cite{Hao2025ShortWinsLong}. 

The two metrics separate. In BLER the long code is steeper, and the two configurations cross near 1.3~dB. In SBERT similarity the difference is qualitative. At 0~dB the long code delivers a similarity of 0.04, a total loss, while the segmented scheme retains 0.70, because most segments still decode correctly and the model repairs a portion of the remaining errors. The choice between one long code and several short ones is a choice of failure mode, complete failure versus graceful degradation. The same design extends to images. On Kodak images tokenized by Emu3, eight LDPC-coded segments with semantic repair reach 26.57~dB PSNR at $-2.7$~dB SNR, while learned deep JSCC baselines with higher channel occupancy remain below 25.4~dB PSNR up to 0~dB SNR.

\vspace{-7pt}
\subsection{Mutual Reinforcement: Iterative Fusion}

\begin{figure}[t]
\centering
\includegraphics{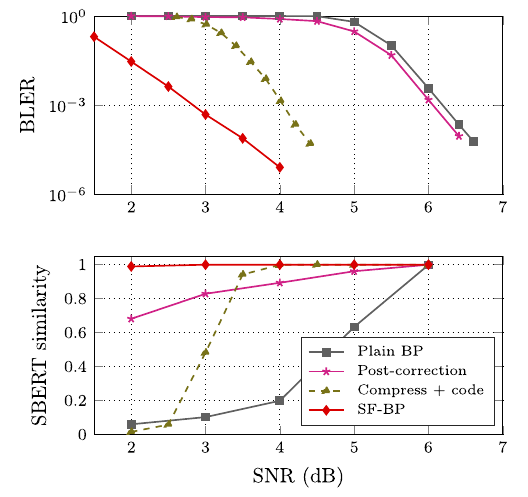}
\caption{Semantic decoding strategies for the $(1248,1040)$ rate-$5/6$ LDPC code with 50 BP iterations. Compress + code applies DEFLATE and a $(1248,832)$ rate-$2/3$ code at equal channel occupancy.}
\label{fig:bler_fusion_comparison}
\end{figure}

Iterative fusion moves the prior inside the decoder. Fig.~\ref{fig:bler_fusion_comparison} reports the results on a $(1248,1040)$ rate-$5/6$ LDPC code with byte-level English text. Semantic-factor BP (SF-BP) converts the byte-level prior into bit-level semantic LLRs and fuses them into belief propagation, while semantic post-correction applies the model once to correct the decoder output. Compress + code compresses each message with DEFLATE and spends the saved rate on a stronger code at equal channel occupancy. Post-correction is a one-shot instance of \emph{coding helps inference}. It barely improves the BLER of plain BP. Without segmentation, a single decoding failure contaminates the entire block, leaving the model too little reliable context to form an accurate prior. SF-BP reaches a BLER of $10^{-3}$ at 2.8~dB, where plain BP requires 6.2~dB. The lower panel reports the SBERT similarity. SF-BP holds the highest similarity at every SNR. In compress + code, a block error corrupts the DEFLATE stream, so the similarity is low at low SNR.

\vspace{-7pt}
\subsection{The Compression Baseline}

The compress + code baseline in Fig.~\ref{fig:bler_fusion_comparison} quantifies the rate cost of Section~\ref{sec:framework}. Compressing each 130-byte message (6.4 bits per character) allows the baseline to use the stronger $(1248,832)$ rate-$2/3$ code, which is worth 2.1~dB under the same decoder. Even so, SF-BP still requires 1.2~dB less SNR at a BLER of $10^{-3}$. The decoding gain from retained redundancy exceeds the rate gain from compression. Two stronger baselines exist. A cross-packet dictionary, as in 3GPP uplink data compression (TS 38.323), reaches about 2.8 bits per character on long text, but a lost packet corrupts the decompression of every packet that follows. Arithmetic coding driven by ByT5 itself reaches 4.8 bits per character on the same sentences, close to the 4 bits per character of BPE tokenization (Section~\ref{sec:framework}). Even against the strongest model-driven compression, tokenization therefore costs little extra rate. However, under arithmetic coding, a single residual bit error desynchronizes the decoder and the whole sentence is lost. In contrast, under SECC with tokenization, residual errors corrupt only the affected tokens, which the model can often repair.

\vspace{-7pt}
\subsection{Latency}

\begin{table}[t]
\centering
\caption{Measured decoding latency. The segmented scheme decodes its blocks in parallel. The long-code and segmented times do not vary with SNR.}
\label{tab:latency}
\renewcommand{\arraystretch}{0.95}%
\begin{tabular}{lcc}
\toprule
\textbf{BCH $\mathbf{(127,64)}$, per codeword} & \textbf{0~dB} & \textbf{3~dB} \\
\midrule
OSD ($m=4$) & 992~ms & 164~ms \\
\semOSD\ & 1990~ms & 281~ms \\
\midrule
\textbf{LDPC $\mathbf{(1248,1040)}$, per sentence} & \textbf{2~dB} & \textbf{6~dB} \\
\midrule
Plain BP & 9.0~ms & 1.25~ms \\
SF-BP & 58.9~ms & 1.43~ms \\
\midrule
\textbf{Segmented transmission, per sentence} & & \\
\midrule
Long code & \multicolumn{2}{c}{1630~ms} \\
Segmented + semantic repair & \multicolumn{2}{c}{63~ms} \\
\bottomrule
\end{tabular}%

\end{table}

Table~\ref{tab:latency} reports the decoding latency, measured on an Intel Core Ultra 7 265 CPU and an NVIDIA GeForce RTX 5090 GPU. All decoders and models run in software, so the absolute times are platform-dependent. For \semOSD\ the prior roughly doubles the decoding time of conventional OSD, and both decrease with SNR as fewer candidates are enumerated. At 2~dB, SF-BP, which updates its semantic LLRs up to six times per codeword, takes 58.9~ms per sentence against 9.0~ms for plain BP, and at 6~dB, where failures are rare, the two receivers differ by less than 0.2~ms. The cost follows the error rate, not the throughput. Segmentation makes the semantic receiver the faster one. With blocks decoded in parallel, the segmented receiver completes a sentence in 63~ms, including semantic repair, against 1630~ms for the long code.

Post-correction and segment repair invoke the model only after a block has failed. In deployed communication systems, a failed block waits for a HARQ retransmission, so the model call competes with the HARQ round-trip delay, not with an additional decoder iteration. The round trip costs milliseconds in cellular links and seconds to minutes in satellite and deep-space links, far above the model call on any platform.

The latency of SECC also decreases with each hardware generation of model inference. Accelerator throughput roughly doubles every two years, and inference software such as quantization and speculative decoding compounds the improvement. The coding gain is independent of the hardware.

\vspace{-6pt}
\section{Open Challenges}

\vspace{-2pt}
\subsection{Unifying Code and Model}

Every scheme in this article couples two separate objects, a code defined over bits and a model defined over tokens. Each exchange between them reduces the model's joint distribution to per-bit LLRs, discarding the dependencies between positions. In one direction, the model is already a code. Plausible sequences (sequences representing meaningful content) form a vanishingly small subset of the full sequence space, just as codewords form a small subset of all binary sequences. The distance properties of this implicit code, however, have never been characterized, so it is exploited only heuristically. In the other direction, the code can be built over tokens instead of bits. Treat each token as one code symbol and add redundancy as parity tokens. Decoder messages and model predictions then share the token domain, both expressed as probability distributions over the token alphabet. The decoder can fuse them by direct multiplication, and no information is lost to bit-level conversion. Unlike JSCC, the structure stays explicit and analyzable. The missing theory is the design rules and distance analysis of such token-domain codes under a model distribution.

\vspace{-7pt}
\subsection{Encoder-Side Semantic Protection}

The demonstrated schemes upgrade only the receiver. The encoder counterpart applies the semantic unequal protection of Section~\ref{sec:secc}. Two problems are open. The first is how to measure importance and match it to the code. Candidate measures include the token probability under the model and the semantic or task quality loss. The measure is continuous, while puncturing patterns and modulation bit levels offer only a few discrete protection levels. The second is how to deliver the protection pattern to the receiver. Explicit signaling costs rate, and deriving the pattern from previously decoded blocks needs no signaling but propagates errors across blocks.

\vspace{-7pt}
\subsection{Performance Limits and Mismatch}

No analysis predicts the coding gain of a given prior. \semOSD\ outperforms the normal approximation bound because the bound assumes uniform information. For text the amount of redundancy is known (Section~\ref{sec:framework}). Speech, images, and video carry redundancy of different structure, and each may need its own characterization. None of them has a finite-blocklength BLER bound that accounts for the redundancy. Two mismatches make the problem harder. First, the receiver's model is not the true source. Mismatched-decoding theory quantifies the loss of a fixed incorrect metric through the generalized mutual information, but no closed form exists when the metric is a learned model. Second, the model conditions on channel-corrupted context, so its prediction is inference from noisy observations of the past. Recent theory gives the minimax risk of this problem and shows that statistically optimal prediction can be computationally infeasible. None of these results yet translate into a BLER benchmark, and that translation is the open work.

\vspace{-7pt}
\subsection{Encryption and Compression in Deployed Stacks}

In current cellular stacks, the Packet Data Convergence Protocol (PDCP) layer compresses the payload, optionally through uplink data compression, and ciphers it before channel coding, so the physical layer carries near-uniform bits. SECC requires residual source structure to be present in these bits. Cellular ciphers XOR the payload with a keystream the receiver can regenerate, and a known keystream bit only flips an LLR sign, so the prior information is preserved through ciphering in principle. No standard interface, however, exposes the keystream to the channel decoder. Compressed bits give the decoder no residual source structure to use. Disabling payload compression preserves the prior at the rate cost of Section~\ref{sec:results}, and tokenized transmission retains redundancy by design, but neither is supported by current standards.

\vspace{-7pt}
\subsection{Model Size versus Coding Gain}

A model call costs milliseconds (Table~\ref{tab:latency}), while latency-critical links budget a fraction of a millisecond per block. Restricting calls to decoding failures lowers the average cost but not the worst case. The key unmeasured quantity is the smallest model that preserves the coding gain. Distillation can target next-symbol prediction of the transmitted source alone, discarding general language ability, and shrink the model. The BLER gain as a function of model size, however, has never been measured. The fastest prior is an $n$-gram table. It costs a table lookup per symbol and is exactly the prior of classical source-controlled decoding \cite{Hagenauer1995}, but it captures only short-range statistics and cannot describe rich or multi-modal sources. The open design problem is the lightweight transformer between the $n$-gram table and the full foundation model, small enough for the latency budget yet keeping the long-range context that SECC relies on.

\vspace{-6pt}
\section{Conclusion}

This article introduced semantic error control coding, which brings foundation-model source priors into standard channel coding. It classified the design methods into three approaches, \emph{inference helps coding}, \emph{coding helps inference}, and \emph{mutual reinforcement}, defined performance metrics from bit-level BLER to meaning-level and task-level fidelity, and supported each approach with experimental examples. Depending on the approach, the model guides the decoder search, repairs failed segments, or exchanges soft information with the decoder, while the code structure stays standard. The demonstrated schemes provide several decibels of coding gain over conventional decoding and improve the meaning-level metrics. Finally, we identified several key open challenges.

\vspace{-4pt}
\bibliographystyle{IEEEtran}
\bibliography{reference/semantic_refs}

\vspace{-3\baselineskip}
\begin{IEEEbiographynophoto}{Chentao Yue}
is an ARC DECRA Fellow with the School of Electrical and Computer Engineering, The University of Sydney, Australia. 
\end{IEEEbiographynophoto}

\vspace{-3\baselineskip}
\begin{IEEEbiographynophoto}{Gaoyang Pang}
is a Postdoctoral Research Associate with the School of Electrical and Computer Engineering, The University of Sydney, Australia. 
\end{IEEEbiographynophoto}

\vspace{-3\baselineskip}
\begin{IEEEbiographynophoto}{Branka Vucetic}
is a Professor with the School of Electrical and Computer Engineering, The University of Sydney, Australia. 
\end{IEEEbiographynophoto}

\vspace{-3\baselineskip}
\begin{IEEEbiographynophoto}{Yonghui Li}
is a Professor with the School of Electrical and Computer Engineering, The University of Sydney, Australia. 
\end{IEEEbiographynophoto}

\end{document}